\documentclass[conference]{IEEEtran}
\usepackage{amsmath}
\usepackage{graphicx}
\usepackage{url}
\usepackage{mcode}
\usepackage{amsmath}
\usepackage{color}
\usepackage{xfrac}
\usepackage{flushend}
\usepackage{subfiles}
\usepackage{float}

\makeatletter
\renewcommand*\env@matrix[1][*\c@MaxMatrixCols c]{%
  \hskip -\arraycolsep
  \let\@ifnextchar\new@ifnextchar
  \array{#1}}
\makeatother

\begin{document}

\title{Mathematical Modelling of  TEAM and VTEAM Memristor Model Using Verilog-A}
\author{
\IEEEauthorblockN{Manoj Mathews
\\}
\IEEEauthorblockA{Dept of Electrical and Computer Engineering\\
Rowan University Glassboro, New Jersey, USA\\
Emails: mathew48@students.rowan.edu}
}
\maketitle

\IEEEpeerreviewmaketitle

\begin{abstract}
Anyone who look into the circuitry world will be familiar with the three fundamental circuit elements - capacitor, resistor and inductor. These circuit elements are defined by the relation between two of the four fundamental circuit variables- current, voltage, charge and flux. However, in 1971, Prof. Leon Chua proposed on the grounds of symmetry that there should be a fourth fundamental circuit element which gives the relation between flux and charge. He named this the memristor, which is the short of memory resistor. This theory was then practically modeled, in May 2008 when the researchers at HP Labs published a paper announcing a model for a physical realization of a memristor. This report mainly focuses on the model of memristor and its applications. The advantages of variable resistance, flexibility, no leakage current, and compatibility with CMOS. The element memristor exhibits different characteristics for different applications which results into the formations of different models of memristor. This paper gives review on different models of memristor. Memristors devices can be used in many applications such as memory, logic, and neuromorphic systems. A computer model of the memristor would be a useful tool to analysis circuit behavior to help in develops application of this memristor as passive circuit element via simulation. In this paper, various Verilog-A model of memristor devices are simulated for sinusoidal inputs and output are verified Various window functions has been used. The circuit analysis of the various memristor models are done

\end{abstract}

\begin{IEEEkeywords}
Memristor, Memristive systems, window function, Verilog-A, TEAM, VTEAM.
\end{IEEEkeywords}

\section{Introduction}
Memristor is a combination of “memory resistor”. It is a passive device with two terminals, where the Magnetic flux is related to the amount of passed electric charge through the device.  Since memristor is not an active element, it cannot store or generate any power. Memristor which is a passive device that provides a functional relation between charge and flux. It is defined as a two-terminal circuit element in which the flux between the two terminals is a function of the amount of electric charge that has passed through the device. A memristor is said to be charge controlled if the relation between flux and charge is expressed as a function of electric charge and it is said to be flux-controlled if the relation between flux and charge is expressed as a function of the flux linkage. In 1971, Leon Chua proposed that there should be a fourth fundamental passive circuit element Memristor. Memristors can be used in quite extensive range of applications. In each application, different characteristics is expected from memristor. For example, in logic and memory applications, an element that can compute, control and store the data after computation is needed. They need to have fast read and write times. The reading mechanism shouldn't change the data while reading. The difference between stored data should be large enough to avoid bad noise margins and have better sensitivity. Also, for storing Boolean data in a memristor, the ratio between Ron and Roff resistances should be high enough. There are other characteristics that are important for memristor applications, such as good scalability, low power consumption and compatibility with conventional CMOS.

\section{Various Memristor Models}
An effective memristive device model needs to satisfy several requirements.It must be sufﬁciently accurate and computationally efﬁcient.It is desirable for the model to be simple, intuitive, and closed-form. It is also preferable for the model to be general, so that it can be tuned to suit different types of memristive devices. Various memristor models\cite{models} which can be used in memristor based circuits are simulated and output are plotted 
 
 \subsection{Linear Ion Drift Model}
In a linear ion drift model is with two resistors connected in series, one resistor represents the high concentration of dopant region with high conductance and the second resistor represents the oxide region with low conductance. In this model a uniform field is also assumed, where the ions have equal average ion mobility µV. this is the original memristor design L.O Chua, which is inaccurate as compared to physical memristor devices.
\begin{figure}[H]
\centering
\includegraphics[width=.5\linewidth]{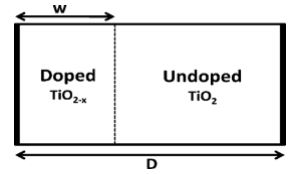}
\caption{Linear Ion Drift Memristive Device Model}
\end{figure}

\begin{figure}[H]
\centering
\includegraphics[width=.5\linewidth]{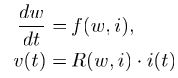}
\end{figure}

\begin{figure}[H]
\centering
\includegraphics[width=.8\linewidth]{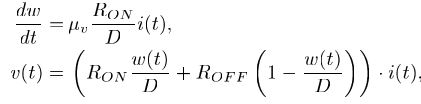}
\end{figure}
 \subsection{Nonlinear Ion Drift Model}
In the nonlinear ion drift model, a voltage-controlled memristor exhibits a nonlinear dependence between the voltage and the internal state derivative is assumed. In nonlinear ion drift model, the state variable w is a normalized parameter within the interval [0, 1]. This model also assumes an asymmetric switching behavior. 

 \subsection{Simmons Tunnel Barrier Model}
In the Simmons tunnel barrier model adopts nonlinear and asymmetric switching behavior due to an exponential dependence of the movement of the ionized dopants, specifically, changes in the internal state variable. In this model, rather than two resistors in series as in the linear drift model, there is a resistor in series with an electron tunnel barrier. The state variable x is the Simmons tunnel barrier width

\subsection{ThrEshold Adaptive Memristor (TEAM) Model}
The TEAM model \cite{team} is a general memristor model. In this model, a current threshold and tunable nonlinear (polynomial) dependence between the current and the derivative of the internal state variable are assumed. The current-voltage relationship can be in a linear or an exponential manner. It is possible to fit the TEAM model to the Simmons tunnel barrier model or to any 

\subsection{VTEAM Model (Voltage Controlled)}
The VTEAM model \cite{vteam} is based on an expression of the derivative of the internal state variable. The VTEAM model combines the advantages of the TEAM model (i.e., simple, general, accurate, and designer friendly) with a threshold voltage rather than a threshold current. The current-voltage relationship of the VTEAM model is undefined and can be freely chosen from any current-voltage characteristics

\section{Window Functions}
To maintain the physical bounds of the device and add nonlinear behavior close to these physical bounds, several window functions are implemented in the Verilog-A model. These window functions are: Jogelkar, Biolek, Prodromakis, TEAM and VTeam.

\section{Need for Threshold Voltage}
The advantages of the TEAM model (i.e., general, simple, and sufficiently accurate) and exhibiting a threshold voltage is desirable. The TEAM model is based on a threshold current. The resistance of the memristor does not change for currents below a certain threshold current.  A memristor with a threshold voltage is more appropriate than a threshold current for certain logic and memory applications.

Similar to the the TEAM model, the VTEAM model is based on an expression of the derivative of the internal state variable. The VTEAM model combines the advantages of the TEAM model (i.e., simple, general, accurate, and designer friendly) with a threshold voltage rather than a threshold current. The current-voltage relationship of the VTEAM model is undefined and can be freely chosen from any current-voltage characteristics. Generally, a voltage-controlled time-invariant memristive device is represented by 
\begin{figure}[H]
\centering
\includegraphics[width=.5\linewidth]{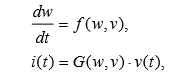}
\end{figure}

 where w is an internal state variable, v(t) is the voltage across the memristive device, i(t) is the current passing through the memristive device, g(w,v) is the device conductance, and t is time. Note that f(w,v) general function of the derivative of the state variable w Specifically, these expressions allow the existence of a threshold voltage. Analogous to the derivative of the state variable in the TEAM model, the derivative of the state variable in the VTEAM model is
 
\begin{figure}[H]
\centering
\includegraphics[width=1\linewidth]{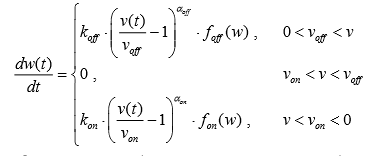}
\end{figure}

\section{Current Voltage Relationships}

The current-voltage relationship is not inherently defined in the VTEAM model. A linear dependence of the resistance and state variable can be achieved, where the current-voltage relationship is
\begin{figure}[H]
\centering
\includegraphics[width=1\linewidth]{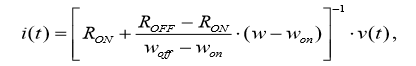}
\end{figure}

An exponential dependence on the state variable can be assumed as in. In this case, the current-voltage relationship is
\begin{figure}[H]
\centering
\includegraphics[width=1\linewidth]{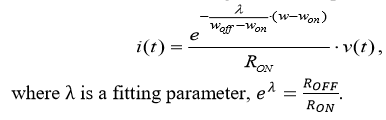}
\end{figure}

\section{Comparison of TEAM and VTEAM with other models}
 
 Memristor models, such as the Yakopcic \cite{Yakopic} and BCM \cite{bcm} models, also exhibit a threshold voltage. Both models, however, operate according to a different state variable mechanism than the VTEAM model. The VTEAM model\cite{vteam} increases the resistance while moving the state variable w towards the boundary woff. In Yakopic \cite{Yakopic} and BCM models\cite{bcm}, however, increasing the state variable decreases the resistance of the device. Although this difference is only based on a different definition and terminology, to accurately compare these models to the VTEAM model, a modification of the original models is required. The I-V relationship of both models is mirrored according to the Vplane and I-plane, i.e., the opposite polarity of the voltage and current are used or, alternatively in circuit terms, the memristor is connected to the opposite polarity. The VTEAM model  and TEAM models IV relationship is shown in Results

\section{Results}
\subsection{Linear Ion Drift Model}

\begin{figure}[H]
\centering
\includegraphics[width=1\linewidth]{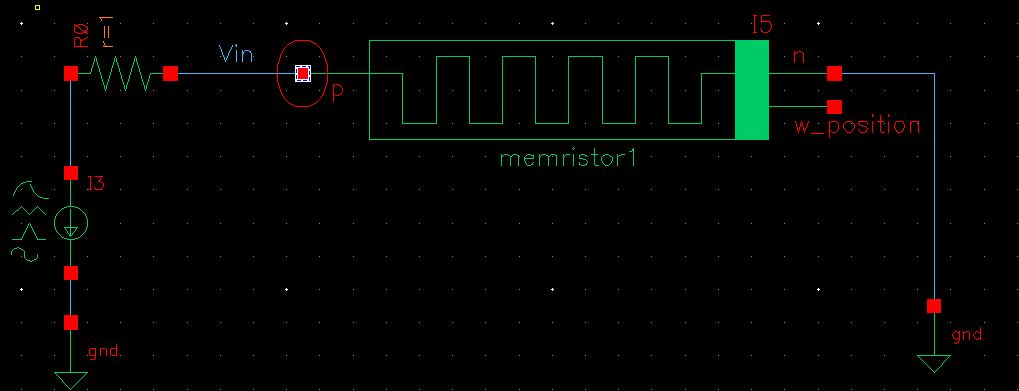}
\caption{Schematic - Linear Linear Ion Drift Model }
\label{Figure1a1b}
\end{figure}

\begin{figure}[H]
\centering
\includegraphics[width=1\linewidth]{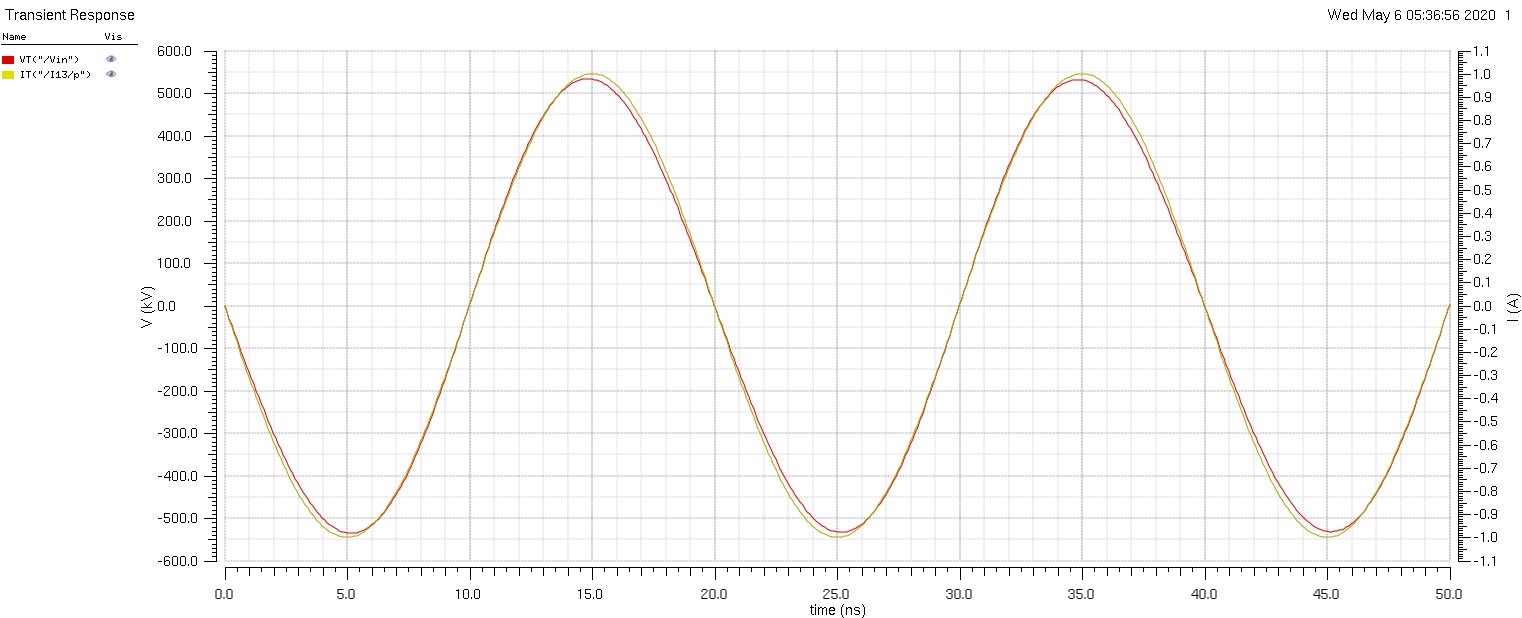}
\caption{ Linear Ion Drift Model Results}
\label{Figure1a1b}
\end{figure}

\begin{figure}[H]
\centering
\includegraphics[width=1\linewidth]{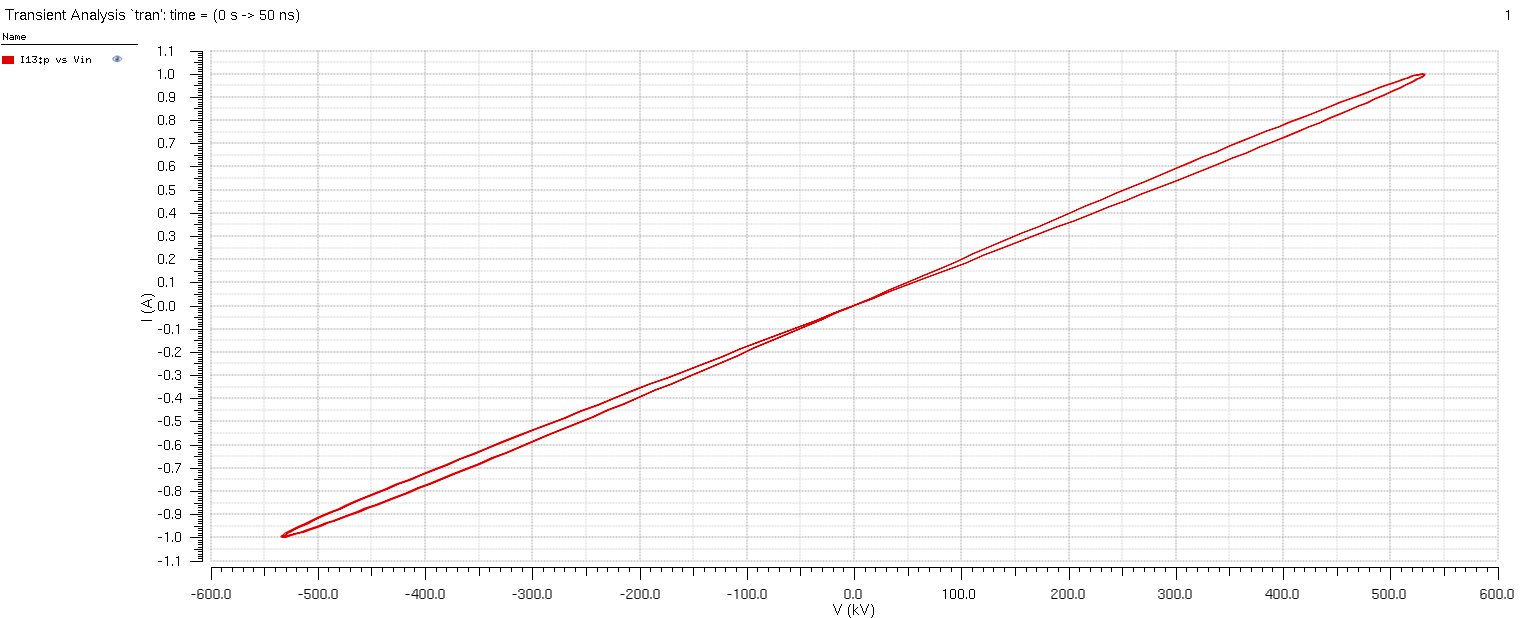}
\caption{Linear Ion Drift model  - VI Characteristics}
\label{Figure1a1b}
\end{figure}

\subsection{Linear Ion Drift Model}
\begin{figure}[H]
\centering
\includegraphics[width=1\linewidth]{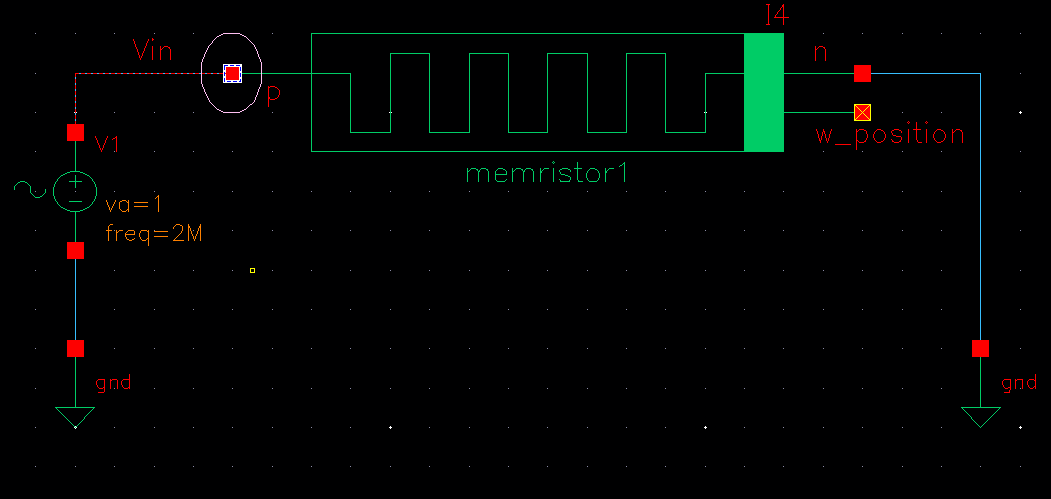}
\caption{Schematic - Nonlinear Ion Drift Model }
\label{Figure1a1b}
\end{figure}

\begin{figure}[H]
\centering
\includegraphics[width=1\linewidth]{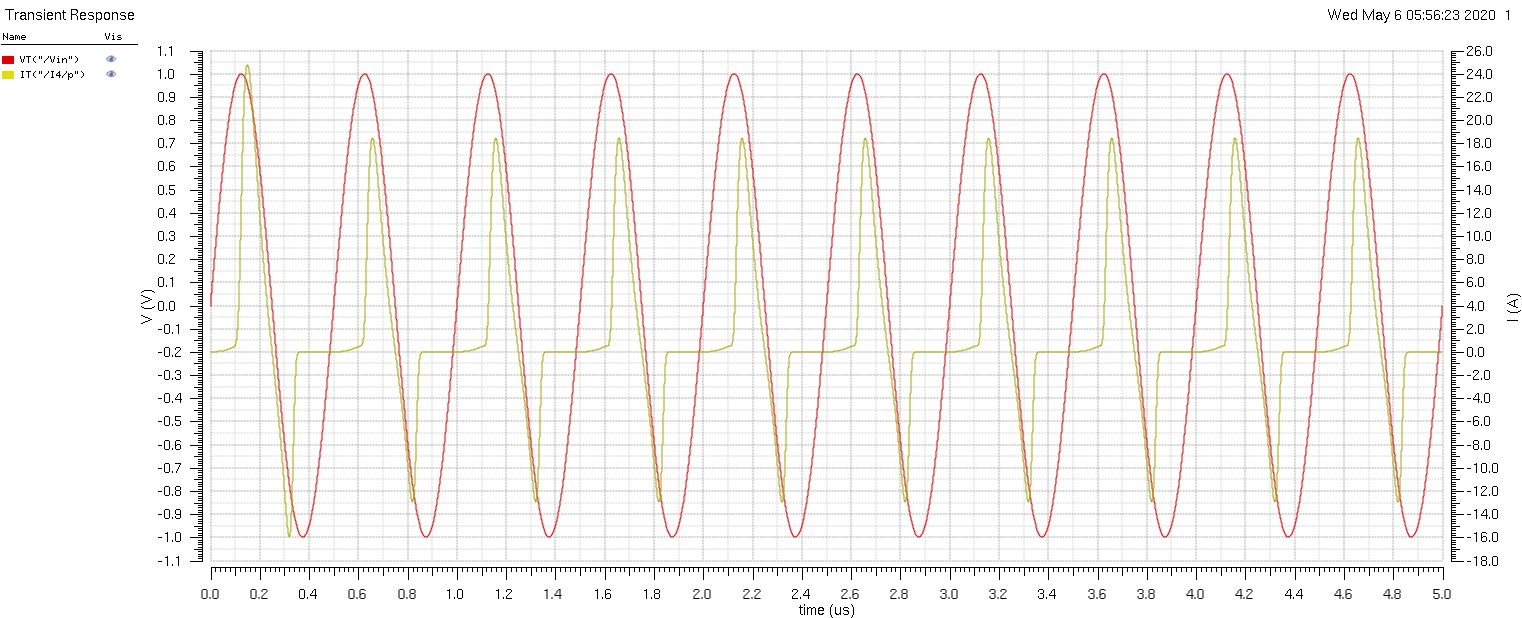}
\caption{Nonlinear Ion Drift Model Results}
\label{Figure1a1b}
\end{figure}

\begin{figure}[H]
\centering
\includegraphics[width=1\linewidth]{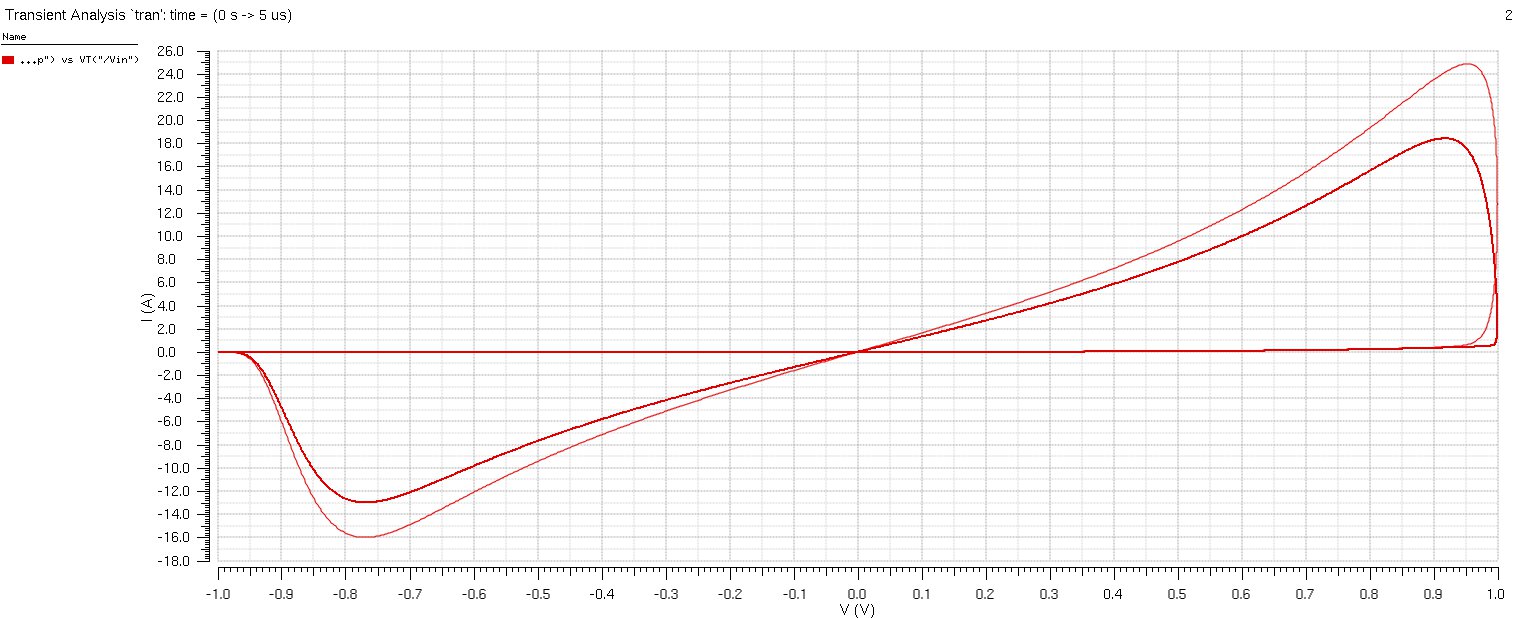}
\caption{Nonlinear Ion Drift model  - VI Characteristics}
\label{Figure1a1b}
\end{figure}

\subsection{Simmons Tunnel Barrier Model}

\begin{figure}[H]
\centering
\includegraphics[width=1\linewidth]{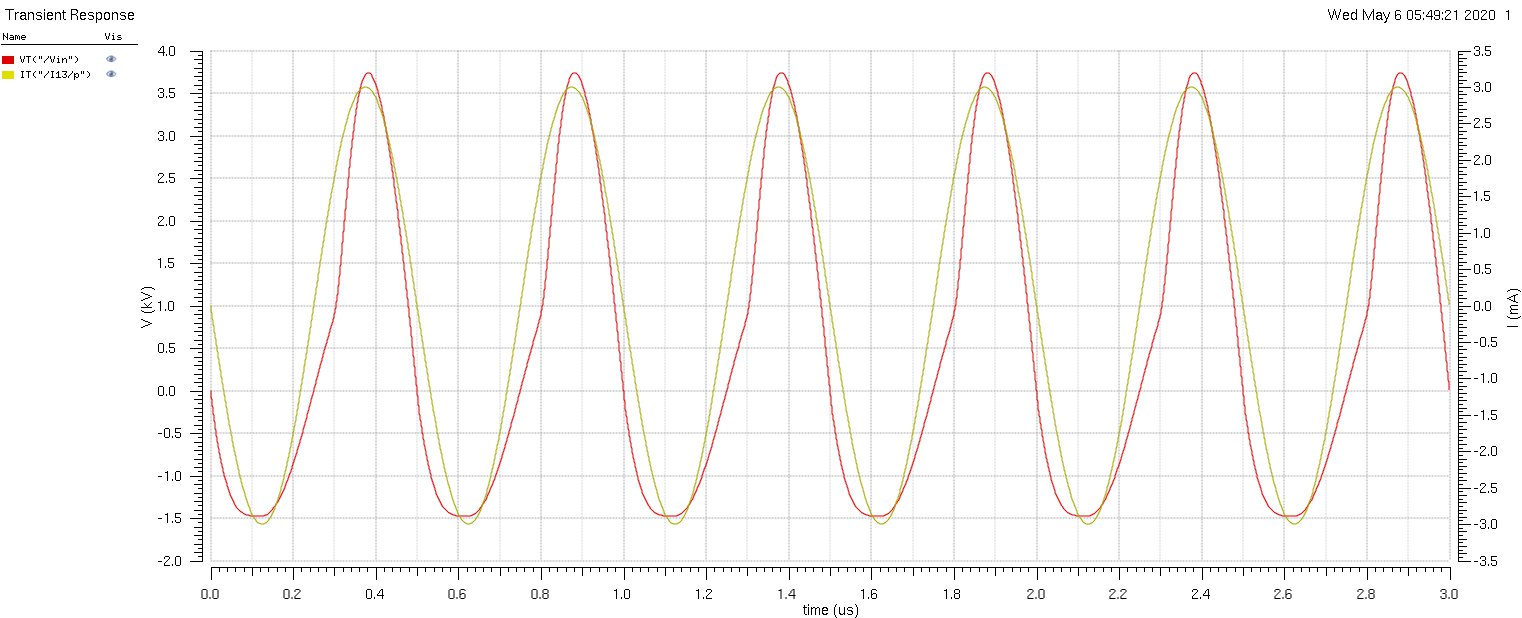}
\caption{Simmons Tunnel Barrier Model Results}
\label{Figure1a1b}
\end{figure}

\begin{figure}[H]
\centering
\includegraphics[width=1\linewidth]{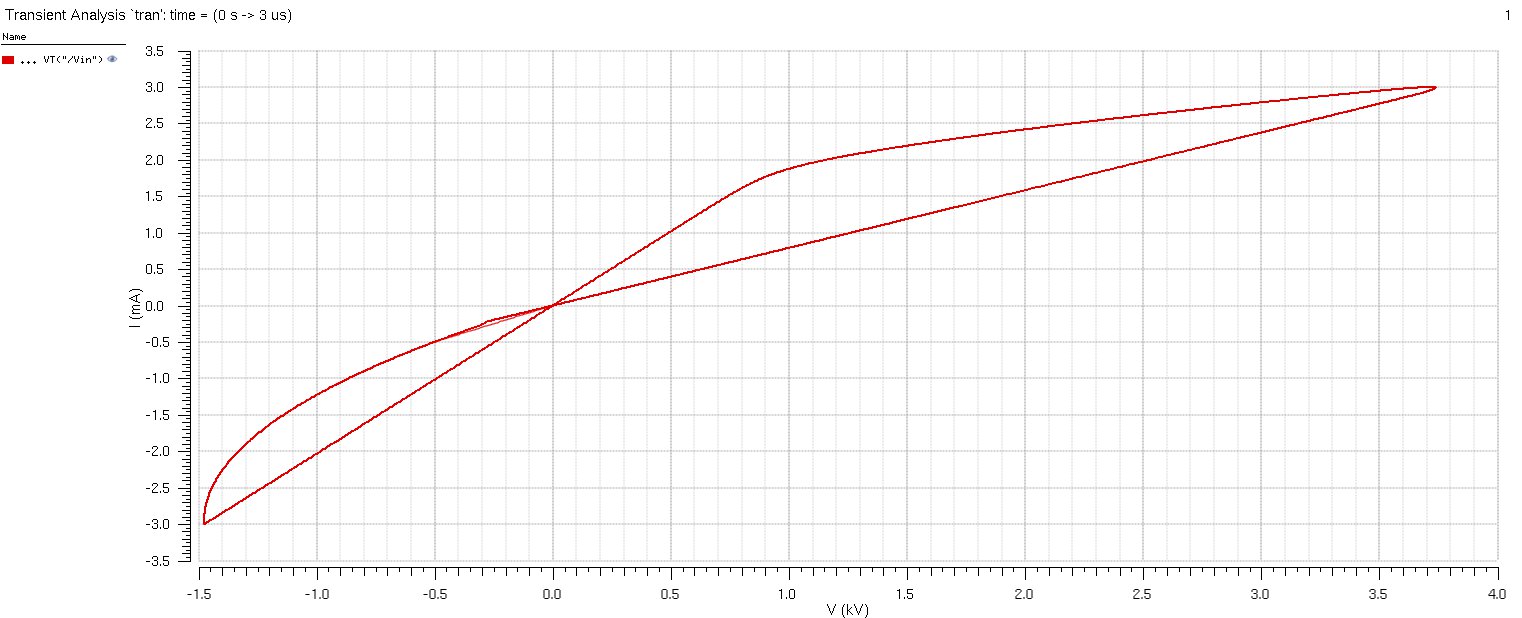}
\caption{Simmons Tunnel Barrier Model - VI Characteristics}
\label{Figure1a1b}
\end{figure}

\subsection{TEAM Model}

\begin{figure}[H]
\centering
\includegraphics[width=1\linewidth]{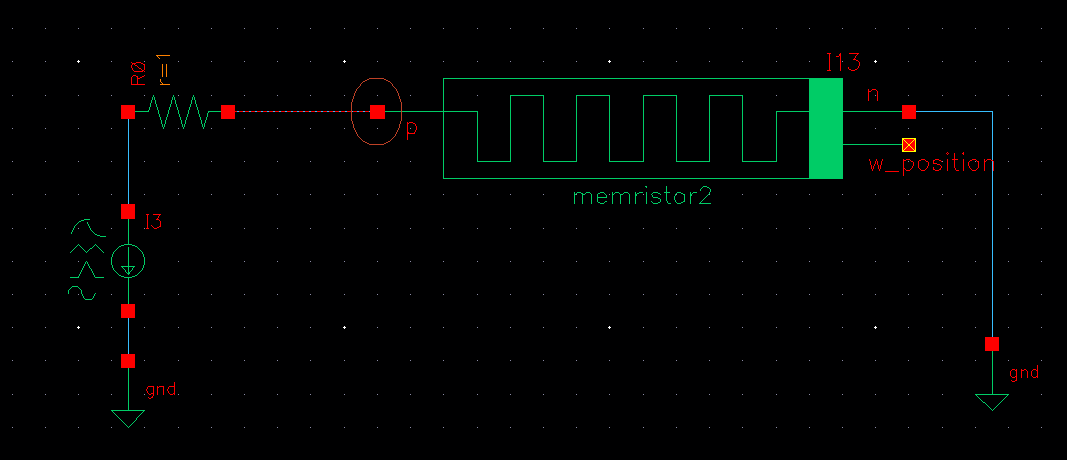}
\caption{Schematic - TEAM Model}
\label{Figure1a1b}
\end{figure}

\begin{figure}[H]
\centering
\includegraphics[width=1\linewidth]{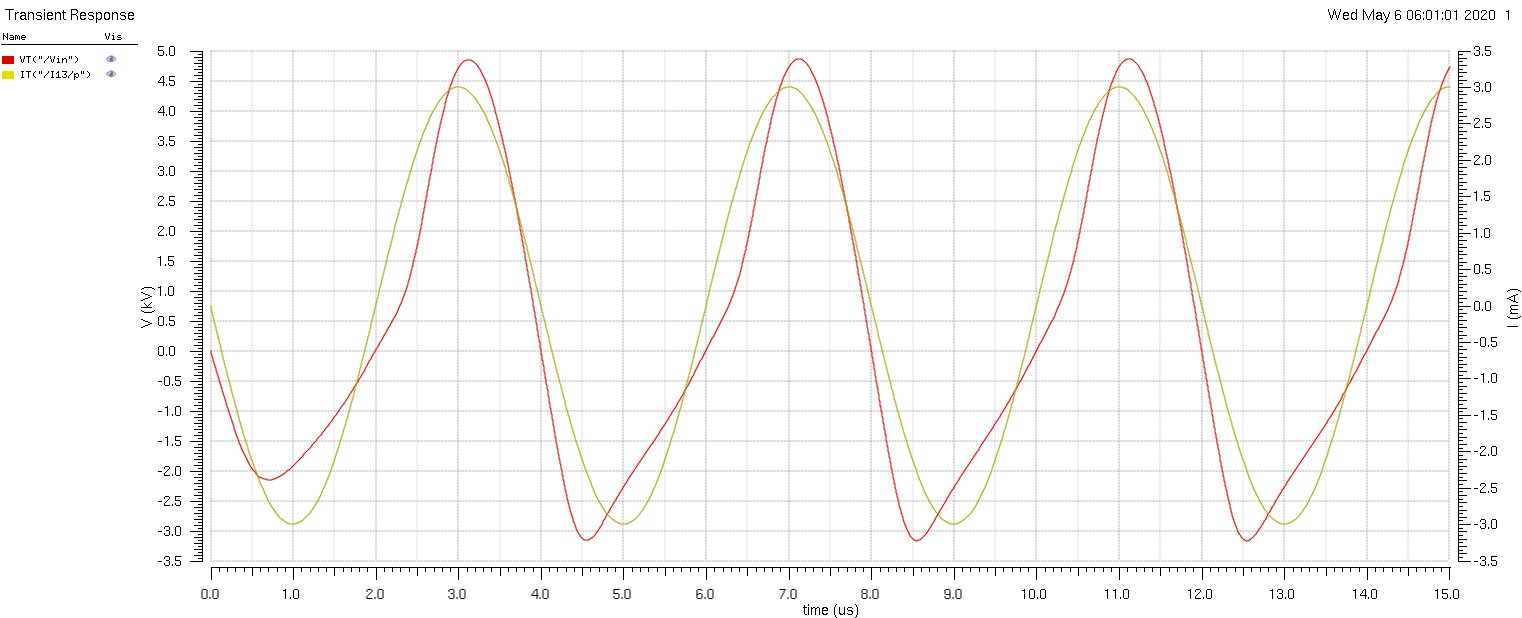}
\caption{TEAM Model Results}
\label{Figure1a1b}
\end{figure}

\begin{figure}[H]
\centering
\includegraphics[width=1\linewidth]{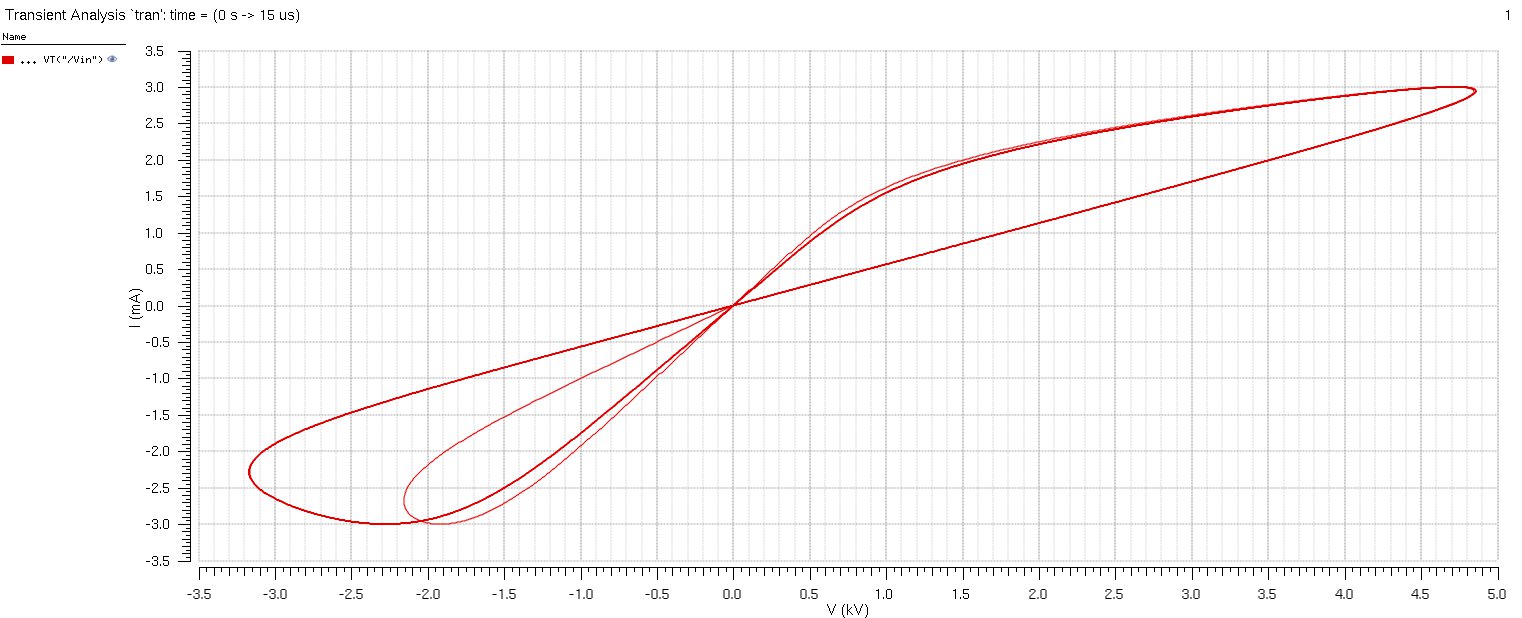}
\caption{TEAM Model - VI Characteristics}
\label{Figure1a1b}
\end{figure}

\subsection{VTEAM Model}
\begin{figure}[H]
\centering
\includegraphics[width=1\linewidth]{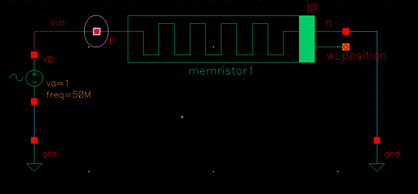}
\caption{Schematic - VTEAM Model}
\label{Figure1a1b}
\end{figure}

\begin{figure}[H]
\centering
\includegraphics[width=1\linewidth]{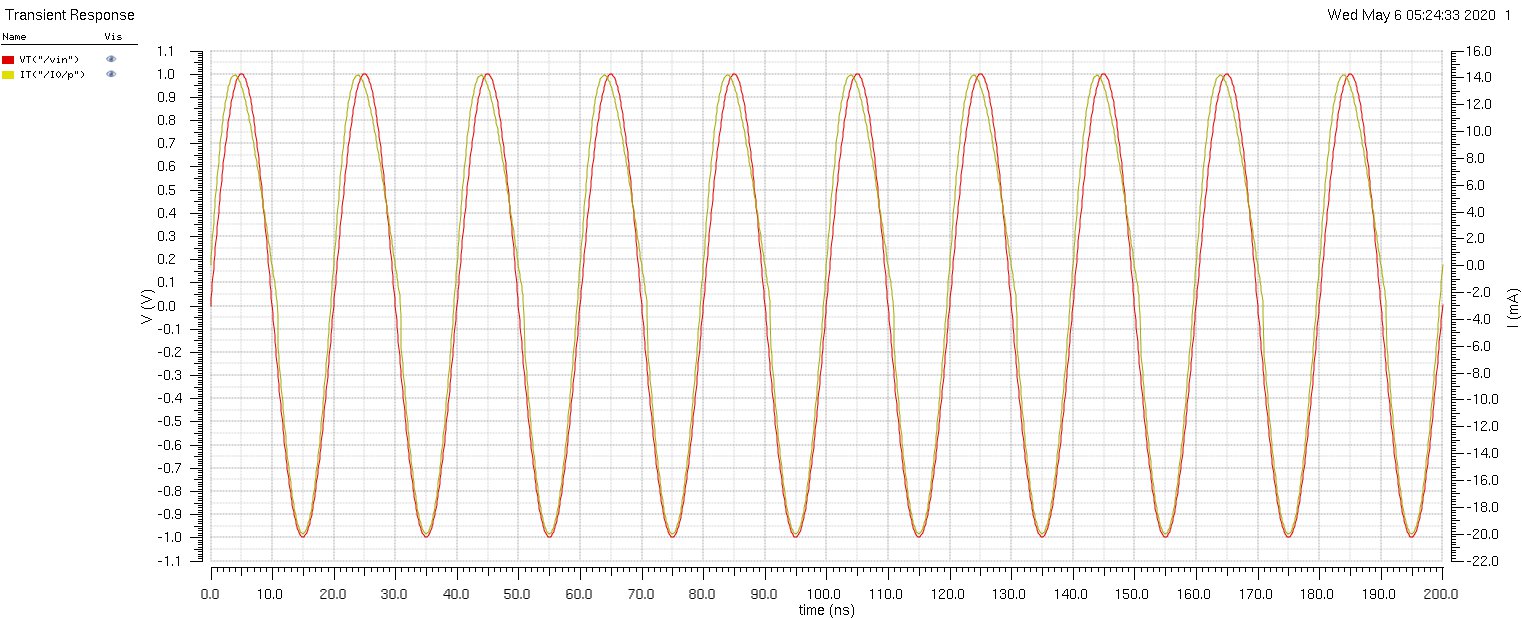}
\caption{VTEAM Model Results}
\label{Figure1a1b}
\end{figure}

\begin{figure}[H]
\centering
\includegraphics[width=1\linewidth]{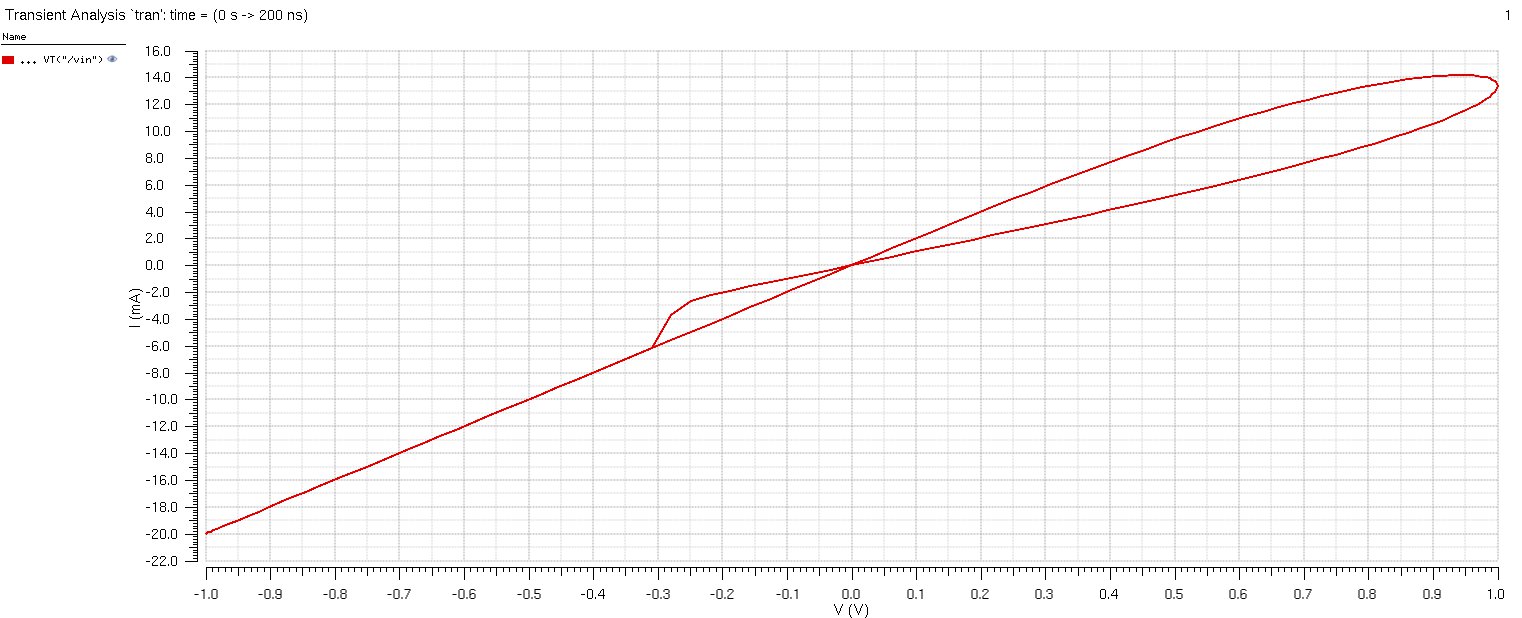}
\caption{VTEAM Model - VI Characteristics}
\label{Figure1a1b}
\end{figure}

\begin{table}[H]
\includegraphics[width=1\linewidth]{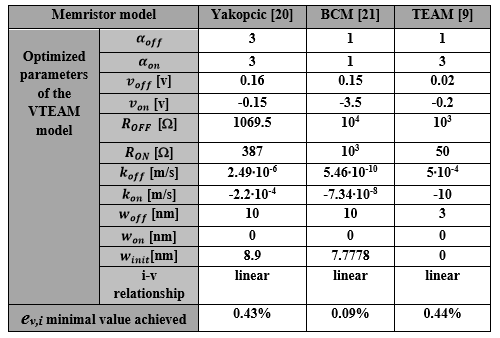}
\caption{Fitting Characteristics of VTEAM Model to other Memristor Models}
\end{table}

\section{Conclusion}
A memristor model that exhibits a threshold voltage is required to accurately characterize physical behavior and to apply to several memory and logic circuits. The VTEAM model\cite{vteam} is  that exhibits a threshold voltage. and has the advantages of the TEAM model (i.e., flexibility, generality, and sufficiently accurate).The simulation output of variours memristor models has been done and compared. Sufficient accuracy of the VTEAM model to experimental data is achieved by tuning the fitting parameters, demonstrating generality and flexibility. The VTEAM model also exhibits sufficient accuracy while fitting to previously proposed memristor models with a threshold voltage. These models lack the generality of the VTEAM model and cannot be fit to experimental data. The VTEAM and TEAM models exhibit, respectively, a threshold voltage and current. These models have been implemented in Verilog-A for SPICE simulations\cite{models}, and can be used to design memristive circuit


\begin{thebibliography}{99}

\bibitem{1} L. O. Chua, “Memristor – The Missing Circuit Element,” IEEE Transactions on Circuit Theory, Vol. 18, No. 5, pp. 507-519, September 1971.
\bibitem{2} L.O. Chua and S.M. Kang, “Memristive Devices and Systems,” Proceedings of the IEEE, Vol. 64, No. 2, pp. 209-223, February 1976.

\bibitem{models}
S. Kvatinsky, K. Talisveyberg, D. Fliter, A. Kolodny, U. C. Weiser and E. G. Friedman, "Models of memristors for SPICE simulations," 2012 IEEE 27th Convention of Electrical and Electronics Engineers in Israel, Eilat, 2012, pp. 1-5, doi: 10.1109/EEEI.2012.6377081.
\bibitem{team}
S. Kvatinsky, E. G. Friedman, A. Kolodny and U. C. Weiser, "TEAM: ThrEshold Adaptive Memristor Model," in IEEE Transactions on Circuits and Systems I: Regular Papers, vol. 60, no. 1, pp. 211-221, Jan. 2013, doi: 10.1109/TCSI.2012.2215714. 
\bibitem{vteam}
S. Kvatinsky, M. Ramadan, E. G. Friedman and A. Kolodny, "VTEAM: A General Model for Voltage-Controlled Memristors," in IEEE Transactions on Circuits and Systems II: Express Briefs, vol. 62, no. 8, pp. 786-790, Aug. 2015, doi: 10.1109/TCSII.2015.2433536.

\bibitem{Yakopic}
C. Yakopcic, et al., "A Memristor Device Model." IEEE Electron Device Letters, Vol. 32, No. 10, pp. 1436-1438, October 2011. 
\bibitem{bcm} 
F. Corinto and A. Ascoli. "A Boundary Condition-Based Approach to the Modeling of Memristor Nanostructures." IEEE Transactions on Circuits and Systems I: Regular Papers, Vol. 59, No. 11, pp. 2713-2726, Nov. 2012
\end{thebibliography}
\end{document}